# Handling massive spatial volumes in time domain simulations using Triggered Cells Method

**Itay Naeh**

*Rafael Advanced Defense Systems LTD*

*P.O.B. 2250, Haifa, Israel*

## Abstract

Simulating wave propagation on discrete grid in time domain requires the description of both the parameters of the media and the values of the wave field at two different time steps within the simulated domain. For most practical cases this operation is very demanding on the system's memory and runtime. In some cases this demands renders the time domain simulation useless. By using the Triggered Cells Method (TCM) one may effectively increase the available memory of a given system by approximately two orders of magnitude in chosen scenarios, and reduce the calculation time by the same factor. The method is most applicable for pulse or bursts propagation, and can achieve outstanding results using single CPU, although it is easily parallelizable for multiple CPU's. This paper will introduce the method using the simple case of the Finite Difference Time Domain implementation of the scalar wave equation, although applicable for most differential equations under any grid type. A hybrid computational engine will be presented, that utilizes the Heterogenic Bulking Method. It will be demonstrated that in most relevant cases, the boundary reflection problem will be resolved using the TCM, without the need to implement sophisticated Phase Matching Layers.

## Introduction

Solving the scalar wave equation using Finite Difference Time Domain (FDTD) methodology[1] provides a robust, straight forward numerical solution for complex boundary conditions problems or propagation in a random media, where analytic solution usually inapplicable. The scalar wave equation is defined by:

$$\frac{1}{c^2_{(x,y,z)}}\frac{d^2\Phi_{(t,x,y,z)}}{dt^2} = \nabla^2\Phi_{(t,x,y,z)} \quad \textbf{(1)}$$

Where $\Phi_{(t,x,y,z)}$ is the wave function at time $t$ and location *(x,y,z)*, $c_{(x,y,z)}$ is the propagation velocity of the wave at position *(x,y,z)* and $\nabla^2$ is the spatial Laplacian operator. The differential formulation of equation **(1)** is therefore:

$$\frac{\Phi_{(t+\Delta t,x,y,z)}+\Phi_{(t-\Delta t,x,y,z)}-2\Phi_{(t,x,y,z)}}{c^2_{(x,y,z)}\Delta t^2}=\frac{\Phi_{(t,x+\Delta x,y,z)}+\Phi_{(t,x-\Delta x,y,z)}-2\Phi_{(t,x,y,z)}}{\Delta x^2}+\frac{\Phi_{(t,x,y+\Delta y,z)}+\Phi_{(t,x,y-\Delta y,z)}-2\Phi_{(t,x,y,z)}}{\Delta y^2}+\frac{\Phi_{(t,x,y,z+\Delta z)}+\Phi_{(t,x,y,z-\Delta z)}-2\Phi_{(t,x,y,z)}}{\Delta z^2} \quad (2)$$

This is the central difference (temporal and spatial) second order accurate formulation. Assuming *dx = dy = dz*, the advancement in time for each grid point:

$$\Phi_{(t+\Delta t,x,y,z)}=2\Phi_{(t,x,y,z)}-\Phi_{(t-\Delta t,x,y,z)}+\left(\frac{c_{(x,y,z)}\cdot\Delta t}{\Delta x}\right)^2\left[\Phi_{(t,x+\Delta x,y,z)}+\Phi_{(t,x-\Delta x,y,z)}+\Phi_{(t,x,y+\Delta y,z)}+\Phi_{(t,x,y-\Delta y,z)}+\Phi_{(t,x,y,z+\Delta z)}+\Phi_{(t,x,y,z-\Delta z)}-6\Phi_{(t,x,y,z)}\right] \quad (3)$$

By updating the next time step at each node one can advance the solution in time. It is important to notice the variables that are needed for the time stepping calculation in the simplest case: $\Phi_{(t,x,y,z)}$, $\Phi_{(t-\Delta t,x,y,z)}$ and $c_{(x,y,z)}$ are the wave field at the current time step, at previous time step and the 3D velocity that describes the media. Usually, at each step, $\Phi_{(t+\Delta t,x,y,z)}$ will override $\Phi_{(t-\Delta t,x,y,z)}$, in a manner that only 2 arrays of $\Phi$ are required.

For performing the desired calculations, the computer's memory must hold $3N^3$ grid points in memory (Two for the wave function and one for the velocity) and perform the update described in equation **(3)** for $N^3$ grid nodes, each time step. Numerical stability requirements determine the spatial spread of grid nodes $\Delta x$ that will produce valid numerical results. A relation that must be kept is the Courant, Friedrich and Levy (CFL) condition[2]:

$$c\cdot\Delta t\cdot\sqrt{\left(\frac{1}{\Delta x}\right)^2+\left(\frac{1}{\Delta y}\right)^2+\left(\frac{1}{\Delta z}\right)^2}\leq 1 \quad \xrightarrow{\Delta x=\Delta y=\Delta z} \quad \frac{c\cdot\Delta t}{\Delta x}\leq\frac{1}{\sqrt{3}} \quad (4)$$

Obviously, both $\Delta x$ and $\Delta t$ must be adequately sampled in a manner that describes the spatial and temporal waveform.

The described restraints severely limit the applicable simulated volume on standard machines in terms of memory and runtime. The Triggered Cells Method (TCM) will demonstrate a sophisticated methodology of selective solving and memory management in order to increase the effective system memory and reduce runtime.

## The Triggered Cells Method

It was stated that the advancement in time is performed at each grid node by **(3)**. The calculation at each node involves its current value, its previous value that is saved in a different array and the values of the neighboring nodes. Let us define this calculation for a single grid node as "Basic calculation", where it will differ for different approximations (central difference second order in this work) and different equations (Scalar wave equation as in this case, Maxwell's equations, elastic wave equations, etc.). The basic calculation is independent of the TCM, and most differential formulations will be applicable for use.

The method takes advantage of the fact that there is a finite travel time for the wave in the simulated domain, and instead of calculating $N^3$ basic calculations in each time step, and allocating at least $3N^3$ values in memory, one can selectively calculate only in regions of the simulated domain where there is a propagating wave and skip the calculation where it is unnecessary. Furthermore, regions that perform no calculations, does not have to be allocated to memory. Thus, by using smart memory management the TCM increases the effective size of the system by allocating only the active regions of the simulated domain. The method will be explained by its steps.

## Cells construction

Let us assume a 3D simulated domain of $N^3$ grid points. The grid will be divided into $M^3$ cells. A cell is a cubic simulated region which contains $\left(\frac{N}{M}\right)^3$ grid nodes. Cells do not contain actual data about the wave function or the media. The cells are actually a spatial division of the simulated domain on a coarser grid. The cells data is stored in a structure which holds information that is required for efficient memory management, as detailed in Table 1. The purpose of all the fields will be explained further on.

| Field | Size | Type | Description |
|---|---|---|---|
| Trig_level | 1 x 1 | Double | The level of the absolute wave field that will cause a cell to be active. Can be set to a constant or relative to the maximal absolute value in the simulated domain |
| Triggered_cells | $M^3$ x 1 | Boolean | Indicates which cell is triggered |
| Triggered_hist | $M^3$ x 1 | Boolean | Indicates which cell was triggered in the previous time step |
| Max_cell_val | $M^3$ x 1 | Double | The maximal absolute value of the wave function in each cell |
| Shutdown_log | $M^3$ x 1 | Integer | The number of time steps remaining for a cell's shutdown process |
| Cell2box | $M^3$ x 1 | Integer | Indicates which box represents a certain cell |
| Box2cell | $M^3$ x 1 | Integer | Indicates which cell is represented in a certain box |
| Linked_cells | $M^3$ x 6 | Integer | Indicates which cells are adjacent to a given cell |
| Diag_cells | $M^3$ x 8 | Integer | Indicates which cells are diagonally adjacent to a given cell |
| Mid_pos | $M^3$ x 3 | Integer | Contains the middle position of each cell |
| Boundary | $M^3$ x 6 | Integer | Contains the boundaries position of each cell |
| homogenic | $M^3$ x 1 | Double | Contains a number that indicates the homogenic velocity within a certain cell. If the cell is heterogenic, contains NaN |

**Table 1 – Structure of the "cells" variable**

When a location that belongs to a certain cell contains relevant information (or is about to), the cell is triggered. When triggered, a sequence commences that allows the computational engine to advance the wave field to the next time step within the given cell. Due to the fact that a cell contains no information about the media or the wave function, a space in memory is allocated in order to perform the calculations on. This allocated space is called a "box".

## Boxes

A box is an allocated space which represents the wave function in a given location of a certain cell. If the wave propagates to a certain location which was not previously initialized, the cell that contains this location is triggered. When triggered, a space in memory is initialized (actually, a box). The size of the box is as the size of the volume described by the cell: $\left(\frac{N}{M}\right)^3$ grid points. The program keeps record that allows fast access for finding which box contains which cell, and vice versa: *Cell2box(i) = j*, and *Box2cell(j) = i*.

When cell *i* becomes untriggered, box *j* is initialized and the variables *Cell2box*, and *Box2cell* stops pointing to each other. When a new cell is activated, it uses the initialized box *j*. For a graphic illustration, see Figure 1.

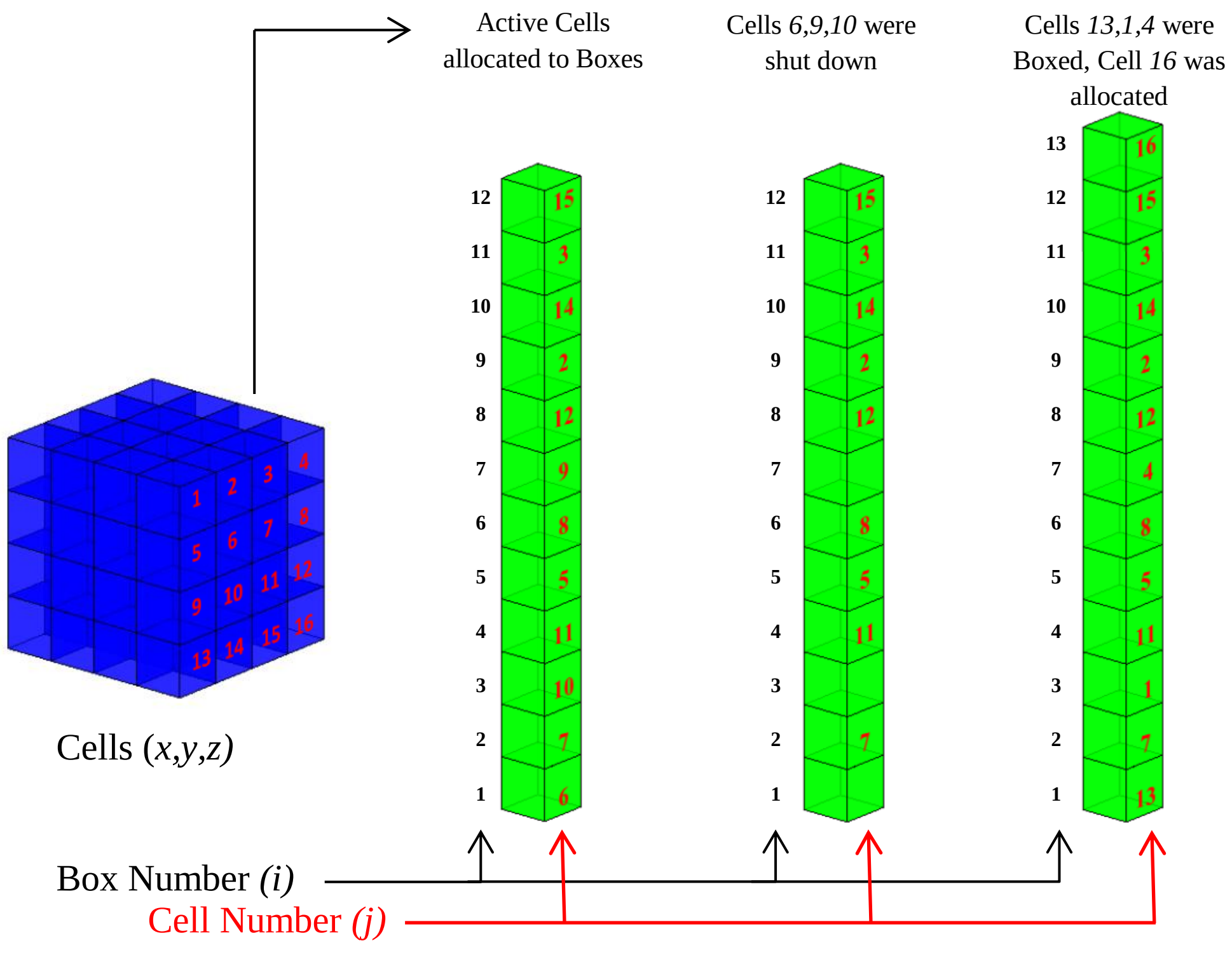

**Figure 1 – The structure of cells and boxes**

## Media

There are several ways to describe the media within the simulated domain. The simplest and fastest is to construct an array of size $N^3$, describing at each spatial location its properties. When memory is limited and time is abundant, one may describe the media by rules. That is – writing a set of conditions that define the structure and values of the media. This method is suitable mainly for describing primitive shapes or analytically defined media, such as the Sparse Spectrum Harmonic Augmentation method[3]. Usually, the "rules" approach carries certain runtime penalty.

The method for describing the media in the TCM is the Heterogenic Bulking Method (HBM), which will be described.

When simulating large volumes with small scattering bodies, it is very easy defining these bodies using rules. Alas, in runtime the operation of implementing these rules each time step is unacceptably slow. In order to save both time and memory, the HBM is used. The first step is done in pre-run and its purpose is to scan the simulated domain, implement all the rules (just once) and divide the cells to two types: Homogenic and heterogenic. If a certain cell *i* contains a single velocity, the variable *homogenic(i)* = *c*, where *c* is that velocity. If within the boundaries of the cell there are variations in velocity, the variable *homogenic(i)* = *NaN*, and a space in memory is allocated to create a 3D array which will hold the velocity structure of the specific cell. An illustrated description is displayed in Figure 2. This process is called "Bulking". Following this methodology, each cell will be either homogenic, and described by a single value, or heterogenic and described by a cell sized bulk in memory. It is important to notice that the slow process of implementing the rules was done only once, and not in every time step. In runtime, if a cell is active, the hybrid computational engine selectively implements different differential operators: When the time stepping function detects a homogenic cell, the coefficient at equation **(3)**: $\left(\frac{\mathbf{c}\cdot\Delta t}{\Delta x}\right)^{2}$ is a scalar, and matrix multiplication is avoided. On the other hand, when the time stepping function detects a heterogenic cell, $\left(\frac{c_{(x,y,z)}\cdot\Delta t}{\Delta x}\right)^{2}$ is taken from the pre-run bulking of the relevant cell and the coefficient is multiplied in a vectorized manner. The HBM demonstrates both fast implementation times and small memory usage, thus benefiting from both approaches described above.

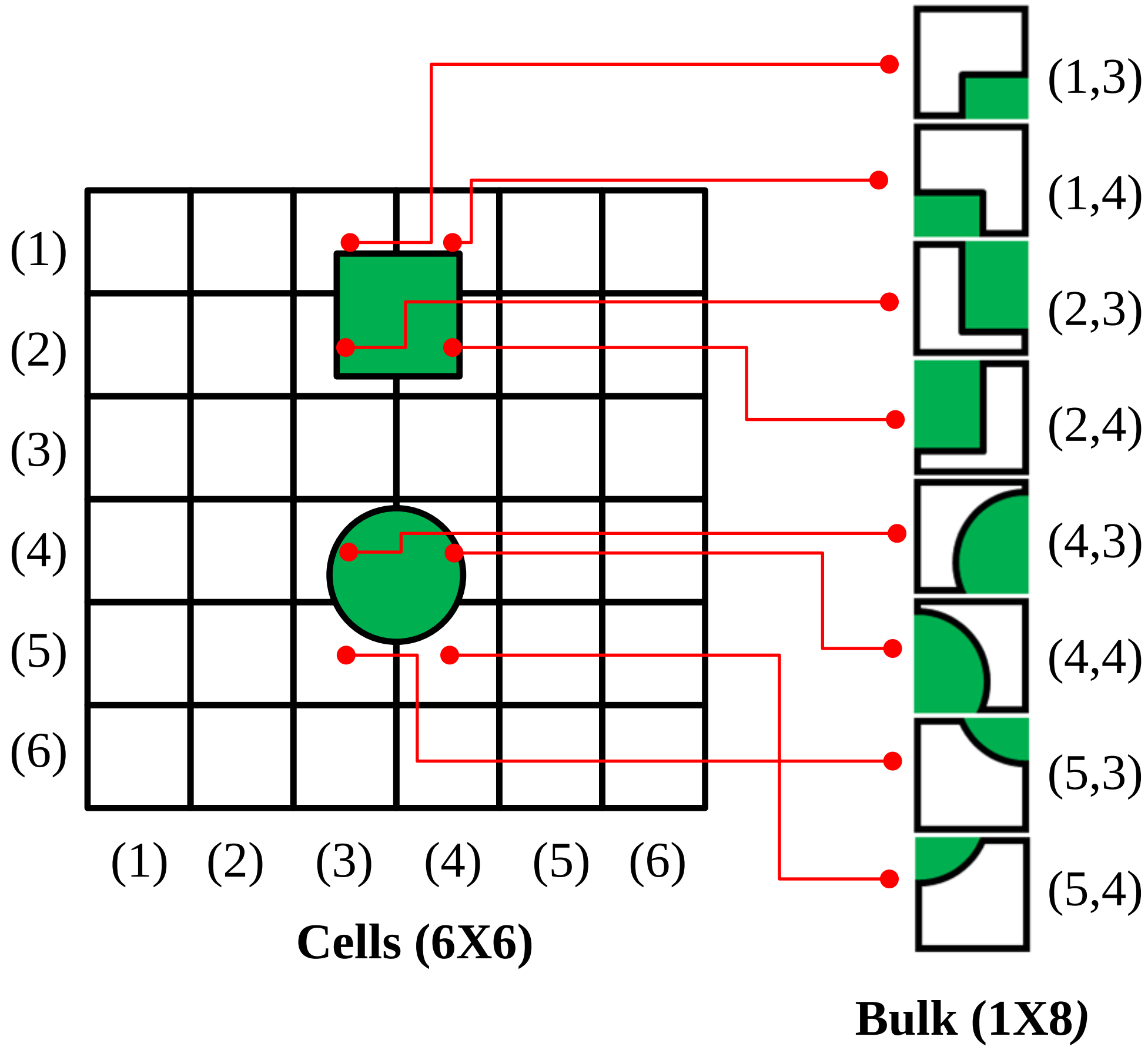

**Figure 2 – Concept of the HBM. A 6X6 cells example of a media containing a sphere and a rectangle with different velocities. If a cell contains several velocities (8 bulked cells in this example), it is tagged as Heterogenic, and bulked explicitly. If a cell contains a single velocity (28 cells in this example), no allocation or bulking occurs, thus saving memory (8 bulked instead of 26) and computation time.**

## Simulation

### *Source insertion*

At the first time steps, the source is being introduced to the simulated domain. Because there is no use introducing a source to an inactive cell, the cell(s) that contain the source have to be manually activated. The manual activation needs to take place until the wave function of the source reaches the triggering threshold. From this moment on, the triggering logic will see to it that the relevant cells will be active. It is important to note that the source is defined in "real" space, and the insertion is being done in the allocated box, which means that there should be a coordinate transformation from "real" space to relative box space.

***Updating active cells***

In order to activate only the relevant cells, there is a process that determines which cells should be active, and which should commence the shutdown process. The activation process includes the following steps:

1. Finding the maximal absolute value of the wave field at each active cell.
2. Registering the current active cells.
3. Keeping cells with maximal absolute value above trigger level, triggered.
4. Triggering nearest neighbors of the triggered cells (See description in Figure 3a).
5. Triggering nearest neighbors in diagonal of the triggered cells (See description in Figure 3c).
6. Initializing shutdown process (will be explained later) for deactivated cells.
7. Unboxing cells that finished the shutdown process.
8. Boxing new cells in vacant boxes, or newly allocated boxes (Described in Figure 1).

The process described above is time consuming (30% of the whole time step is dedicated to this management process), therefore it is not carried every time step. There is a tradeoff between accuracy (updating each time step) and performance (updating every large number of time steps). The optimal updating interval is chosen by the requirement that the wave will not travel the distance of one cell without updating the active cells list. Therefore, the step update interval $\boldsymbol{n}$ can be defined as:

$$\boldsymbol{n} = \frac{\boldsymbol{\Delta x} \cdot (\boldsymbol{Cell\ Length})}{\boldsymbol{2} \cdot \boldsymbol{c} \cdot \boldsymbol{\Delta t}} \tag{5}$$

It is a good practice to place factor of 2 in the denominator in order to be sure that evanescent head waves do not propagate to inactive cells.

The triggering threshold in the TCM can be constant or adaptive. After the cell updating process, the maximal absolute value of the wave function in each cell is known. The trigger level can be set as a certain fraction of the maximal value of the entire domain, therefore compensating for the geometrical spreading and loss of amplitude.

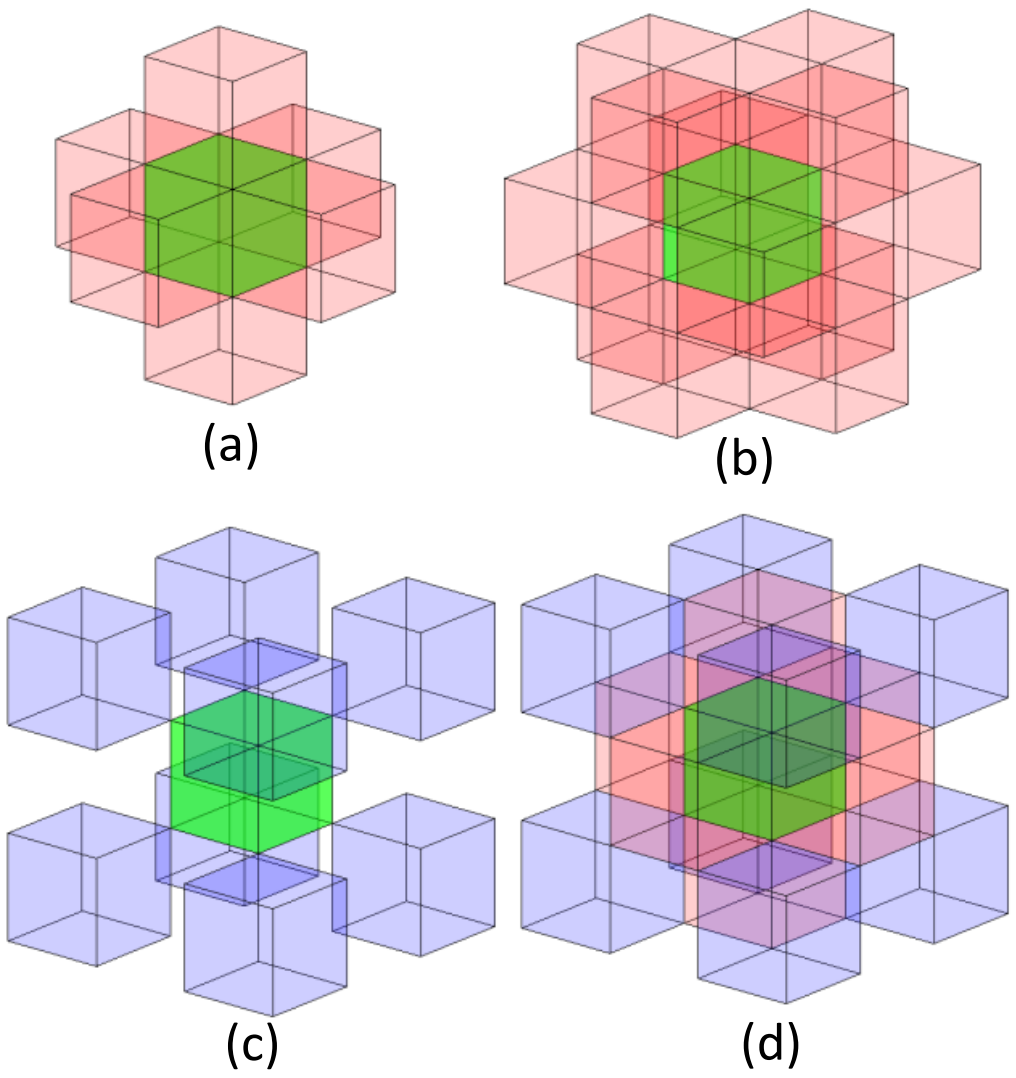

**Figure 3 – Illustration of nearest cells (a), close diagonal cells (b), far diagonal cells (c),**

**and a combination of nearest cells and far diagonal cells (d)**

### *Time stepping*

After determining which cells are active, the computational engine acts on each cell and updates its values to the next time step. When doing so, one must keep in mind that the computational engine actually acts on boxes and not on cells. That is, the boundaries of a certain box are not connected to the boundaries of its neighbors. The meaning of this is that the "basic calculation" as was defines in equation **(3)** can't update boundary nodes without padding the box from all sides by the boundaries of its relevant neighbors. The implication of this demand is that there is a need to keep record of the previous time step's boundaries of all active cells.

### *Shut-down process*

When a wave passes a cell and the wave field drops below the triggering value, the cell becomes untriggered. In this case, the cell needs to be "turned off" in order to save calculation time and memory. Due to numerical stability constraints, a cell can't be turned off in a single time step. Even for very low amplitudes, this would be like introducing high spatial and temporal frequencies to the system, which by definition can't be supported in a given $\Delta t, \Delta x$ discretization. Therefore a cell shut-down process should take place.

In order to avoid divergence, several approached may be applied, and all keep the same principle – stopping the calculation at a given location at a given time step, without introducing unsupported spatial and temporal frequencies. The solution may lie within time domain or frequency domain formalism. Let's assume a cell which is turned off, who is neighboring a cell which stays triggered. In time domain, one can slowly decrease the amplitude of the wave function in the decaying cell, thus not introducing unwanted temporal frequencies. Due to the fact that there is a finite propagation velocity to the wave, this corresponds to not introducing high spatial frequencies as well. This process should be applied through several time steps when the cell remains active. The user should determine the decay factor α at each time step and the number of time steps *n* for the decay process. therefore, the total attenuation factor A at the end of the process will be:

$$\boldsymbol{A = \alpha^n} \tag{6}$$

In frequency domain, the process may be invoked in one time step. It involves transforming the wave function of each neighboring cells to frequency domain using a 3D Fourier transform, filtering the undesired high frequencies by applying a low-pass filter (LPF) and inverse transforming the filtered wave function. In order to keep the spectral characteristics of the original wave form, spatial oversampling is advised in order to separate the introduced numerical noise from the original spectral content.

$$\boldsymbol{\widetilde{\Phi} = F^{-1}\left[F(\Phi) \cdot LPF\right]} \tag{7}$$

## Numerical stability

For comparing the numerical accuracy and noise introduced by the TCM, the same scenario was solved by both the TCM and regular FDTD method. The scenario for checking these issues was chosen to be the scattering from composite shapes. The shape was chosen to be a sphere adjacent to a cube as seen in Figure 4. The wave velocity inside these shapes is set to zero, and the wave is scalar (such as an acoustic wave). The velocity of the media itself was set to *340 [m/sec]*. The source was chosen to be a directional Gaussian beam transmitted from the boundary of the simulated domain. The temporal source shape was set to be a bi-polar pulse defined by the derivative of a Gaussian with $\sigma = 0.4$ *[sec]*. The simulation parameters were: *dt=1e-3 [sec], dx=1 [m]*.

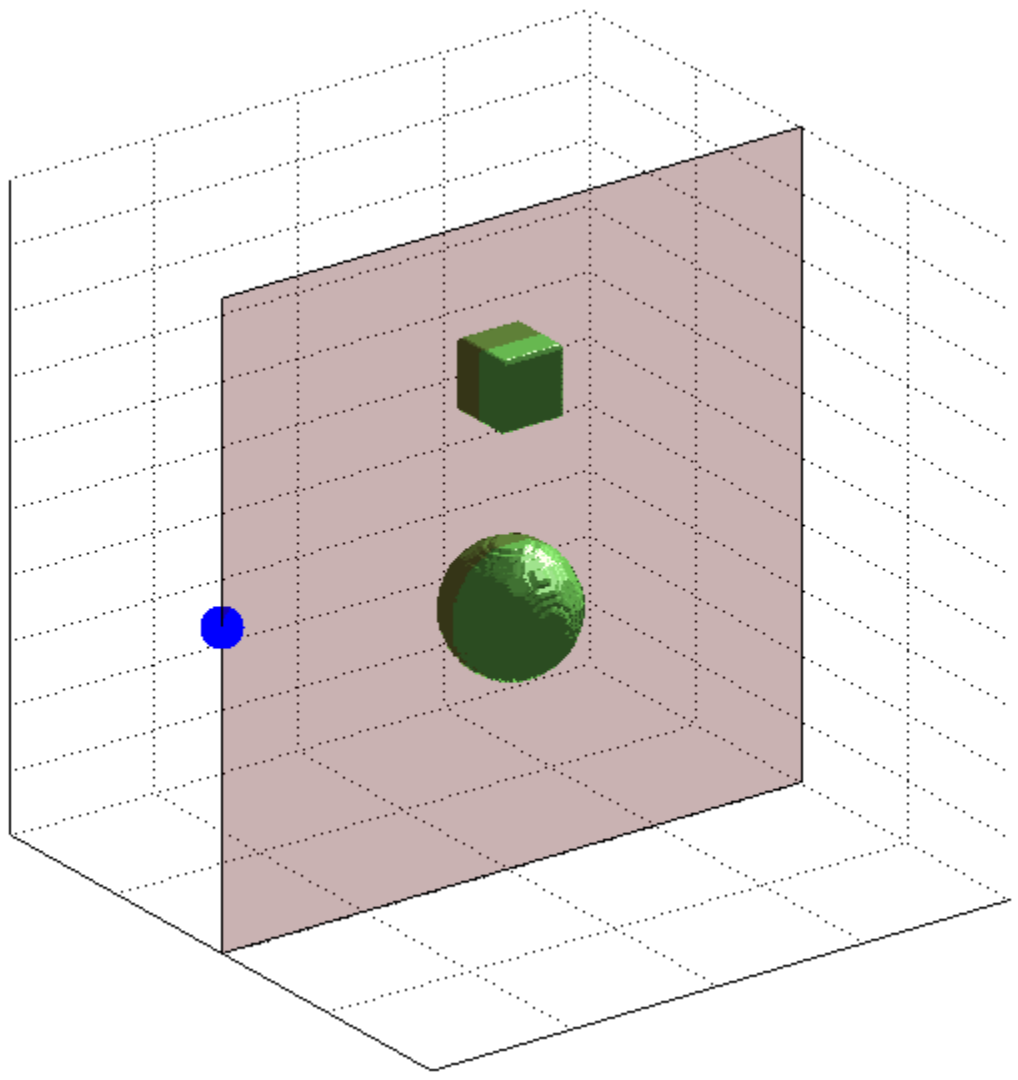

**Figure 4 – The benchmarking Scenario. The blue dot represents the location of the source. The intersecting plane describes the plane in which later comparisons will be made**

Figure 5 – shows the wave function at the central slice of the 3D simulated domain. Figure 6 – shows the comparison of the wave function along *y=500 [m]* of the two methods. The power spectrum of the profile in figure 6 is shown in figure 7. It is noticeable that the PSD similarity lasts for more than 3 orders of magnitude below the main spectral content of the wave function. Within the scope of the given parameter set, the signal fidelity is well kept to the $4^{th}$ order scattering, as seen in figures 8 and 9. The user may trade free memory and runtime to achieve better fidelity by lowering the trigger level and tweaking the decay factors α and *n* in:

$$A = \alpha^n \tag{6}$$

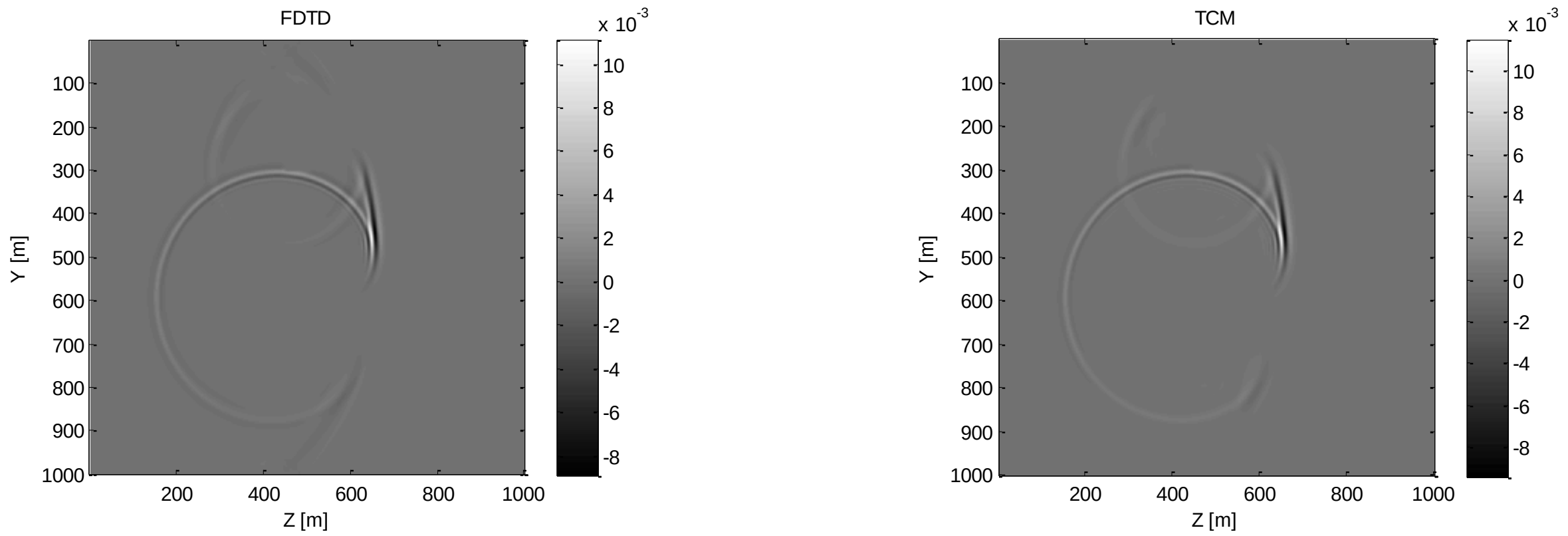

**Figure 5 – Snapshot comparison between the same scenario calculated with the regular FDTD method (Left) and with the TCM (Right)**

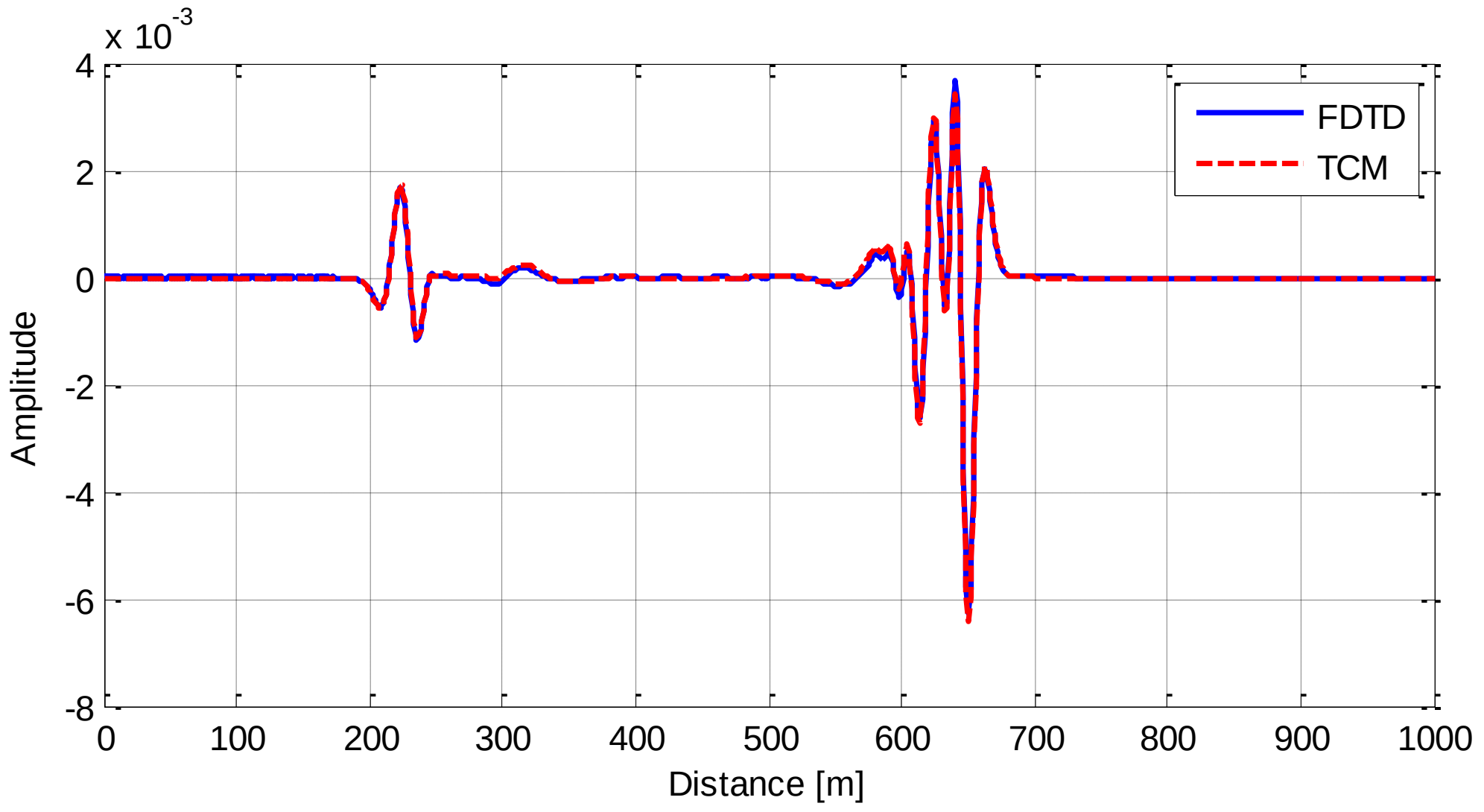

**Figure 6 – The signal cross section of Figure 5, along the line *y = 500***

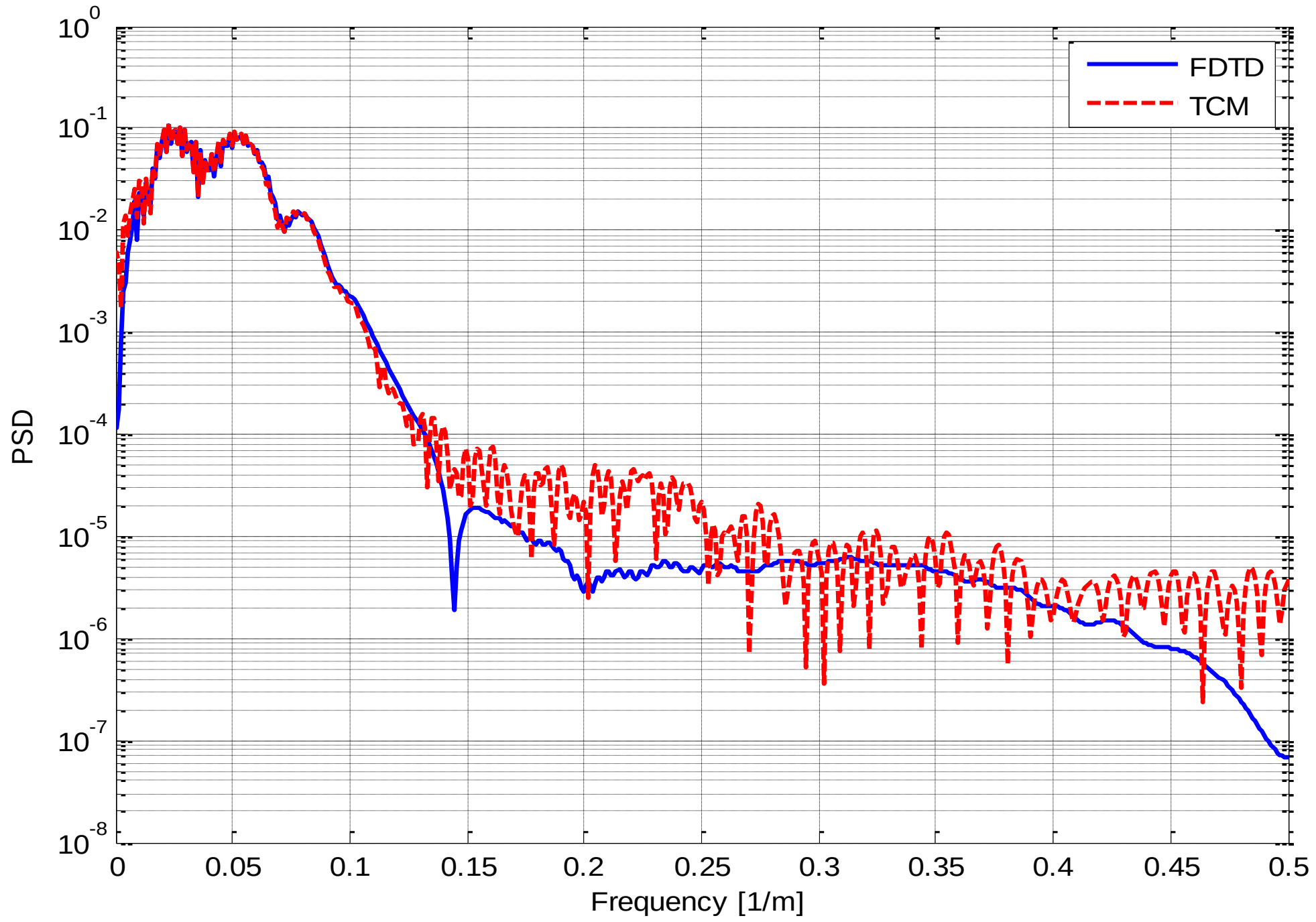

**Figure 7 – The power spectrum of the signal in Figure 6**

## Boundary reflection elimination

Reflections from the boundaries of the computational grid are a common problem in the area of numerical simulations. There are many methods for dealing with this problem. Most of them are from the large family of the Perfectly Matched Layer methods[4,5]. The purpose of these methods is to simulate an infinite space within a finite grid near the boundaries of the domain, thus eliminating back reflection. With the TCM there is no real boundary, for in suitable cases there is enough memory to describe the outward propagating wave front. Therefore, no reflection occurs. Just as no PML implementation is perfect for each impinging wave in terms of frequency and incident angle, the TCM itself also implements a shutdown process which incorporates a gradual decay with time, unlike the spatial decay of conventional PML methods. Both approaches introduce numerical noise to the wave function. Choosing the correct parameters diminishes these noises, but usually applying the optimal parameters directly implies longer runtime.

In some cases, the whole scenario fits a regular computational grid without memory limitations. But, if one would like to examine a scenario of high order reflections, the initial wave will bounce back from the edge of the simulated domain, even with a sophisticated PML, and will obscure the high order scattering. In order to demonstrate the usage of the TCM as an effective PML, an extremely large scenario of $4000^3$ grid points was set, and the $4^{th}$ order reflection between the two shapes could be seen without artificial boundary conditions (actually, the wave travelled only 1000 grid nodes, so it had another *3000* grid points to travel before the reflection would interfere with the region of interest). It is important to note that such grid implemented with the TCM is 64 times larger than the maximal grid size feasible for regular FDTD method on the same machine.

Figures 8 and 9 describes the region of interest of the scenario (zoomed in, displaying a movie on a *4000 X 4000* grid is visually challenging). The $4^{th}$ order reflection is clearly visible.

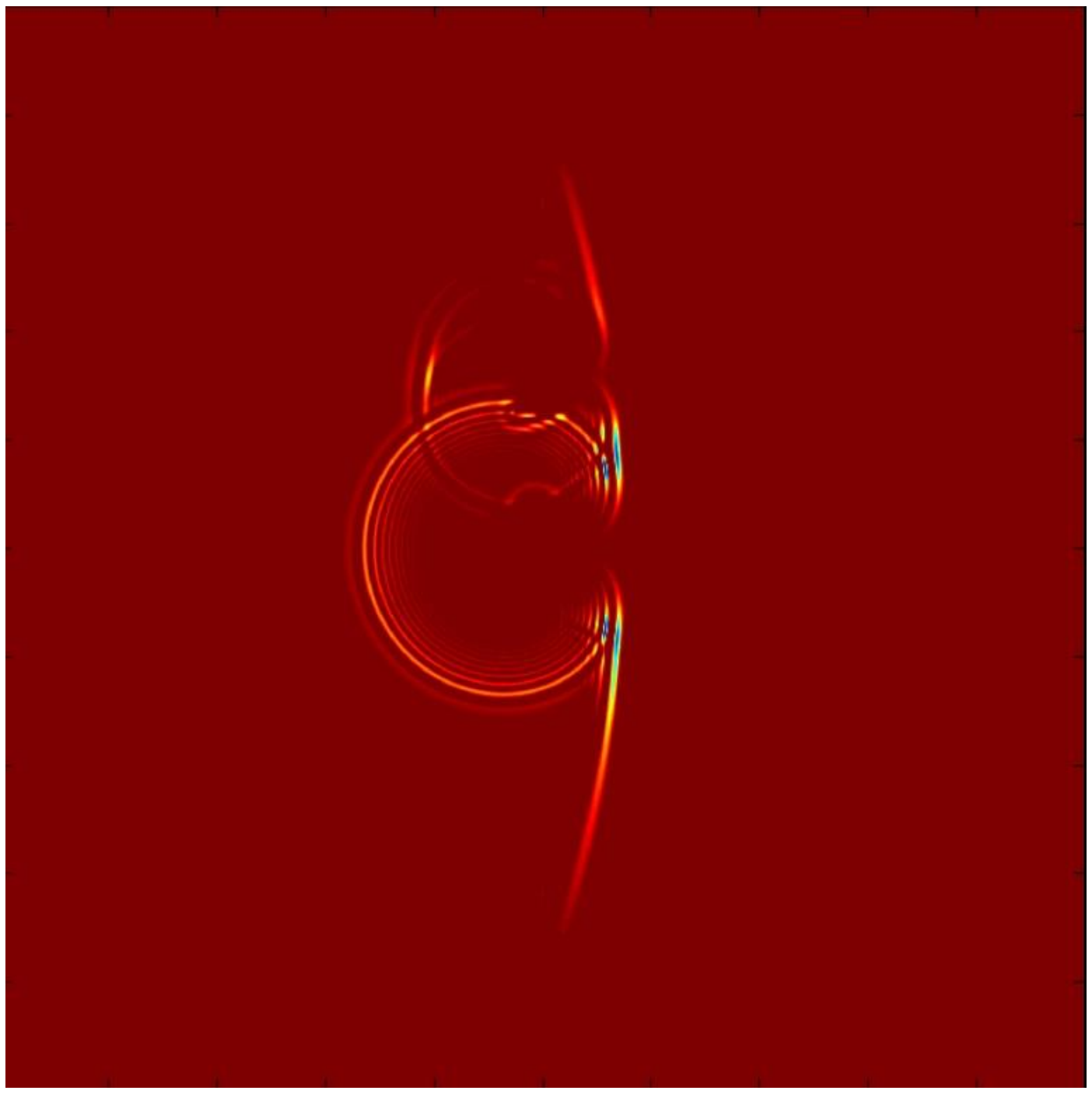

**Figure 8 - A 2000 X 2000 region of interest out of 4000 X 4000 grid points describing the wave propagating and interacting with the shapes**

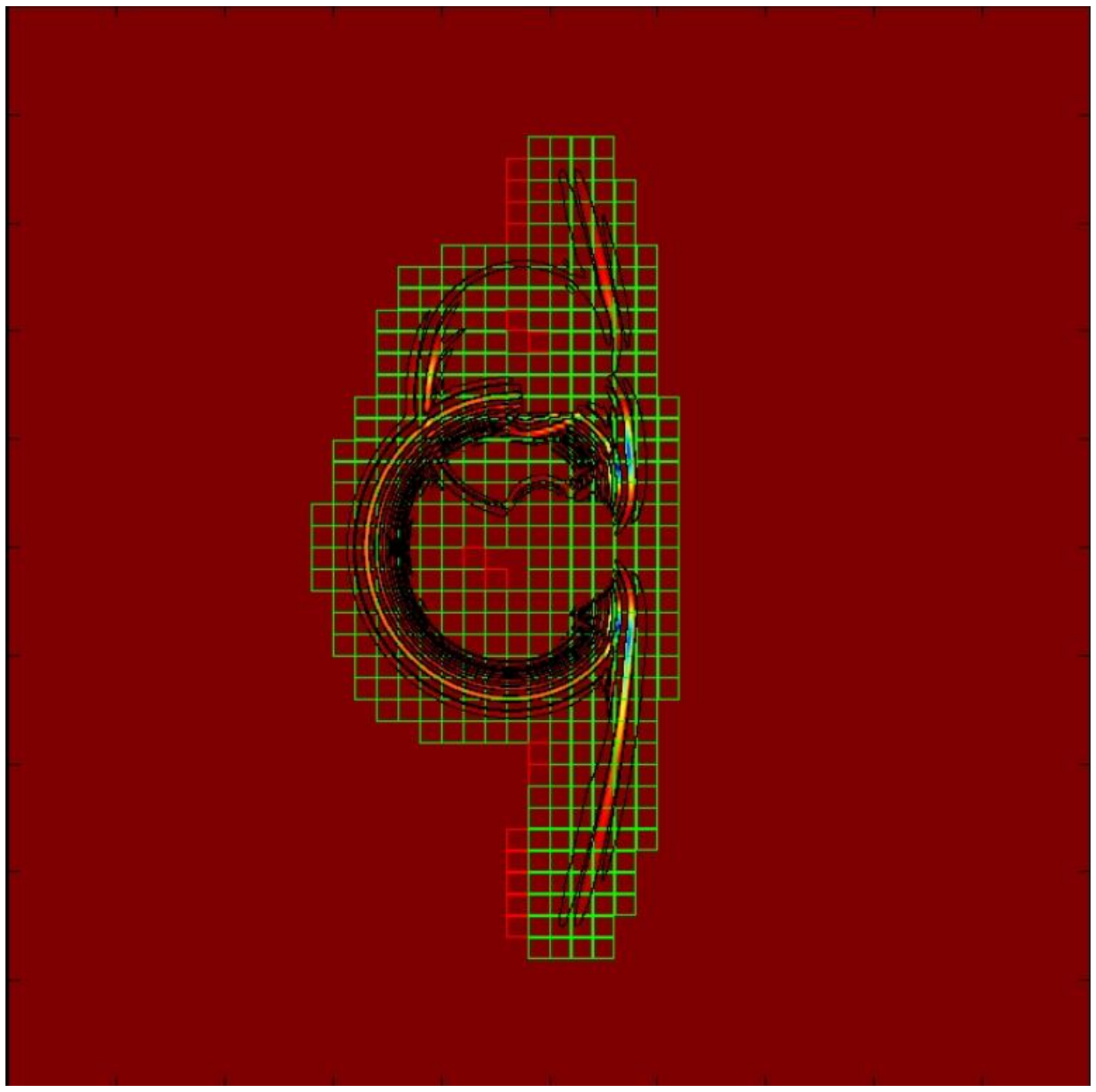

**Figure 9 - As in Figure 8, with the additional display of the active cells at the cross section displayed in green. The red cells are cells that are in their shut down process. The thin black contour around the wave fronts marks the current trigger level for activating new cells.**

## Results

The resulting speedup and the effective memory increase depend on many factors. Therefore, the scattering scenario described earlier will be used for benchmarking. Although the cross section of a sphere and a cube is analytically known separately (under restricting conditions), the high order near field scattering must be solved numerically, as is the scattering from arbitrary shapes and sources. The simulated domain was set to be *2000 X 2000 X 2000* computational nodes, almost an order of magnitude above what the specific computer could manage when not using the TCM, but regular FDTD.

Figure 10 shows the memory allocated for the described scenario. When using standard FDTD method, the allocated space should be proportional to $N^3$, and with the TCM there is a factor of *0.0407* in the required allocated memory, meaning an increase of *25* times in the effective memory of a given system. Figure 11 shows the time it took to run each time step with the TCM, for each type of operation (Basic calculation – blue, Decay – green, triggers updating - black).

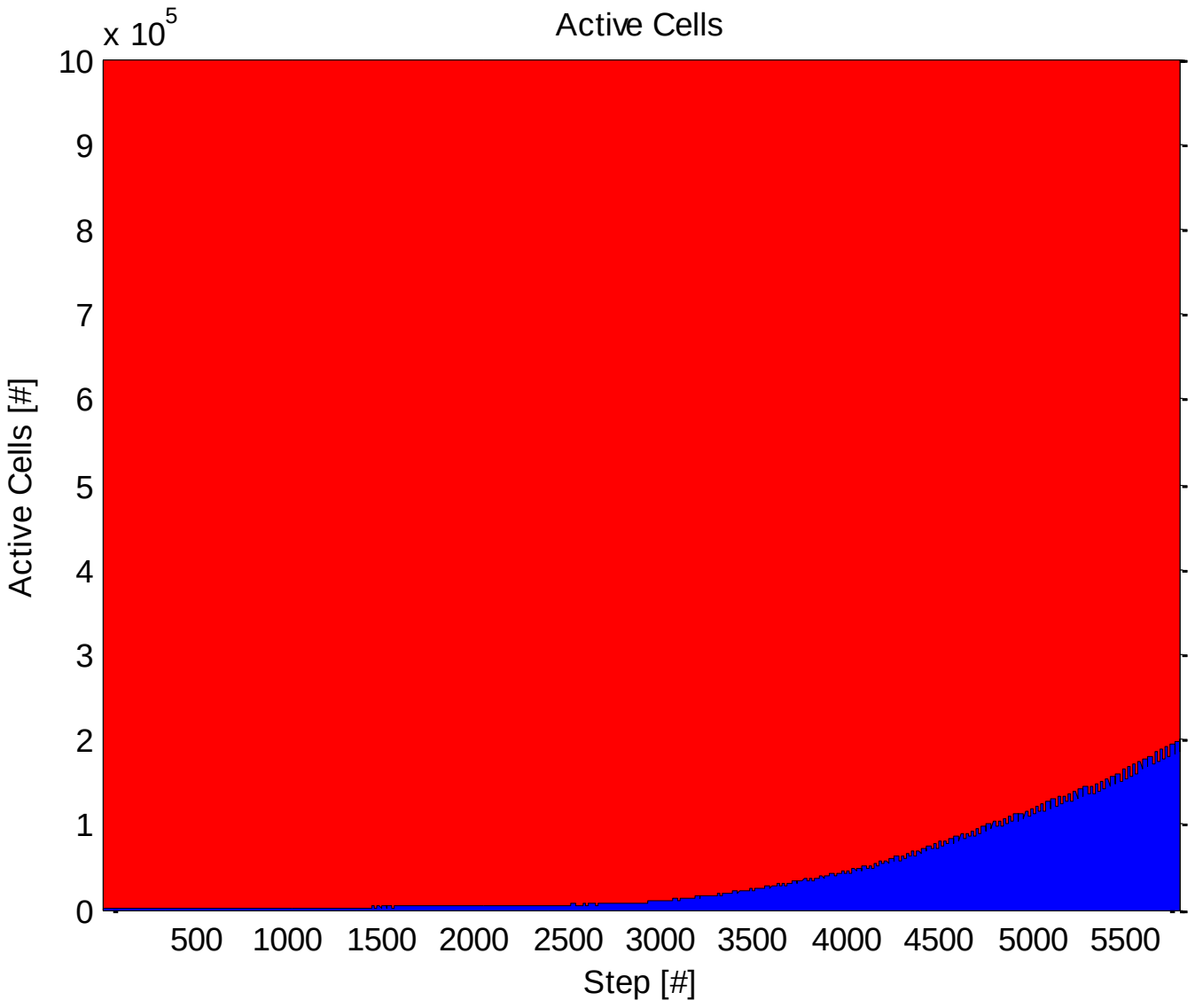

**Figure 10 – The relative allocated memory of the TCM (Blue) and the regular FDTD method (red)**

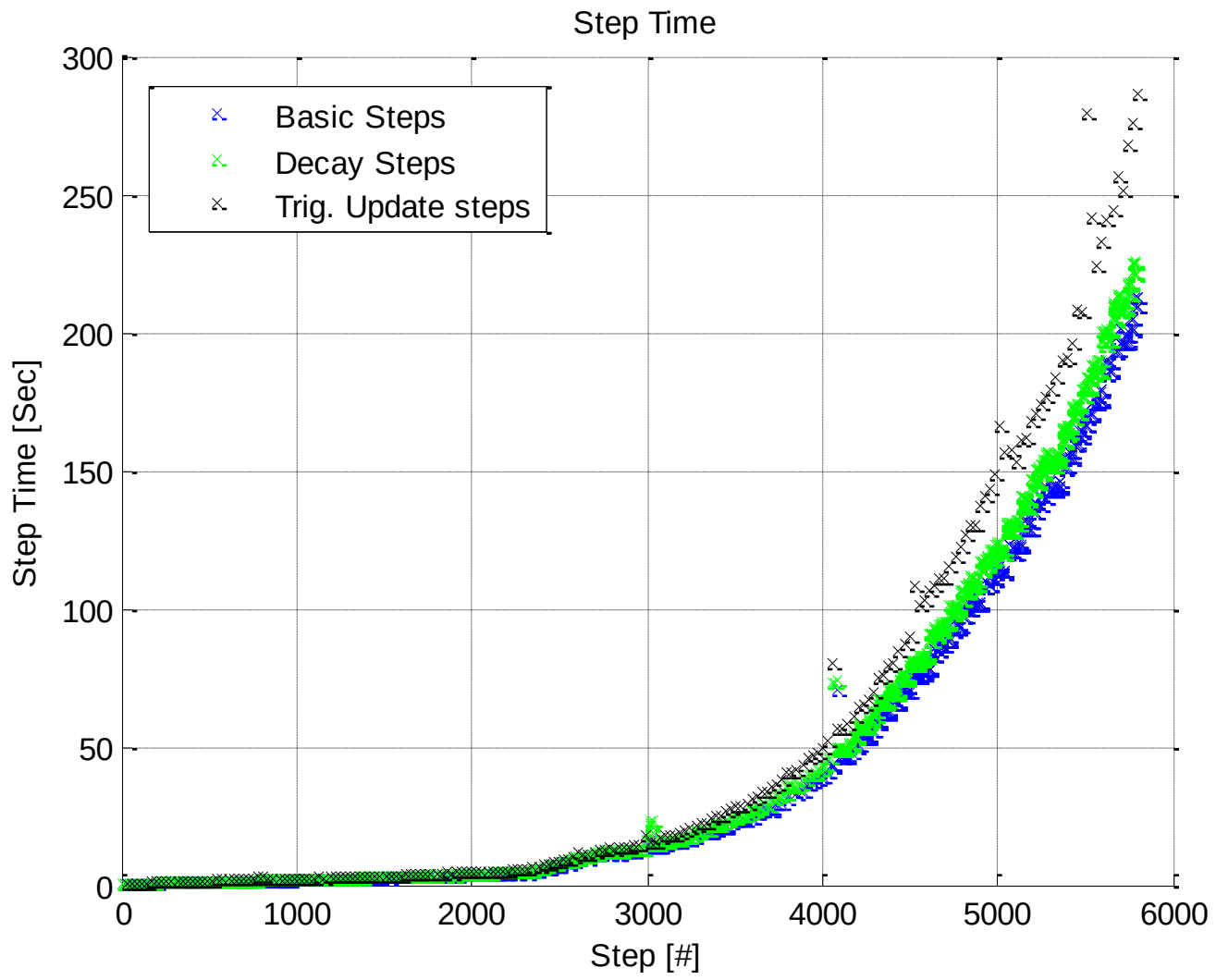

**Figure 11 – Step time for each simulation step**

Comparing the TCM with regular FDTD algorithm is difficult, for the main advantage of the TCM emerges in scenarios with large grids, which can't be allocated for regular FDTD methods due to memory constraints. Nevertheless, the average time step in the given TCM scenario on a grid of 8E9 elements was *43 [sec]*, where with the FDTD method with a grid of 1E9 elements, each time step took *70 [sec]*. Assuming linearity in large grids, one may extrapolate the speedup to be of the order of factor *13* for equal size grids.

The effectiveness of the memory management scheme is demonstrated in figure 12. There are three visible lines: The blue represents the steps which contains basic calculations. The green describes the steps which invoked the decay process (20 steps). The black describes the steps in which the trigger updates took place (every 29 steps). The fact that the trend is quite linear demonstrates that the management overhead is small and uncorrelated to the number of managed cells.

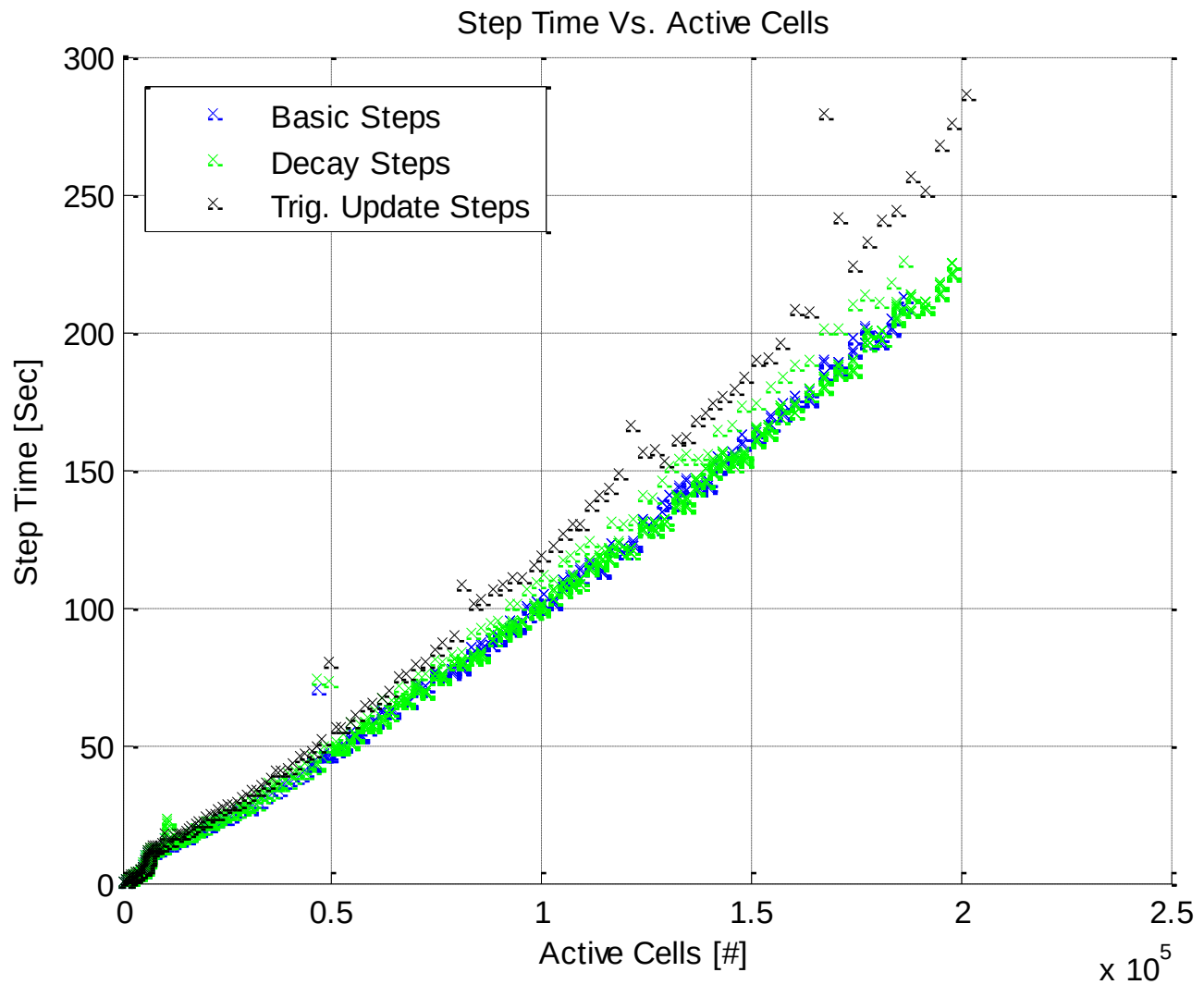

**Figure 12 – Step time as a function of the number of active cells**

In total, the scenario of *2000*$^3$ grid points displayed a speedup factor of *13*, and the effective memory of the given system was increased by a factor of *25*. The scenario of the *4000*$^3$ grid points displayed a speedup factor of *26* and the effective memory increase by a factor of *192*, for simulating 4$^{th}$ order reflections, as described in figure 13.

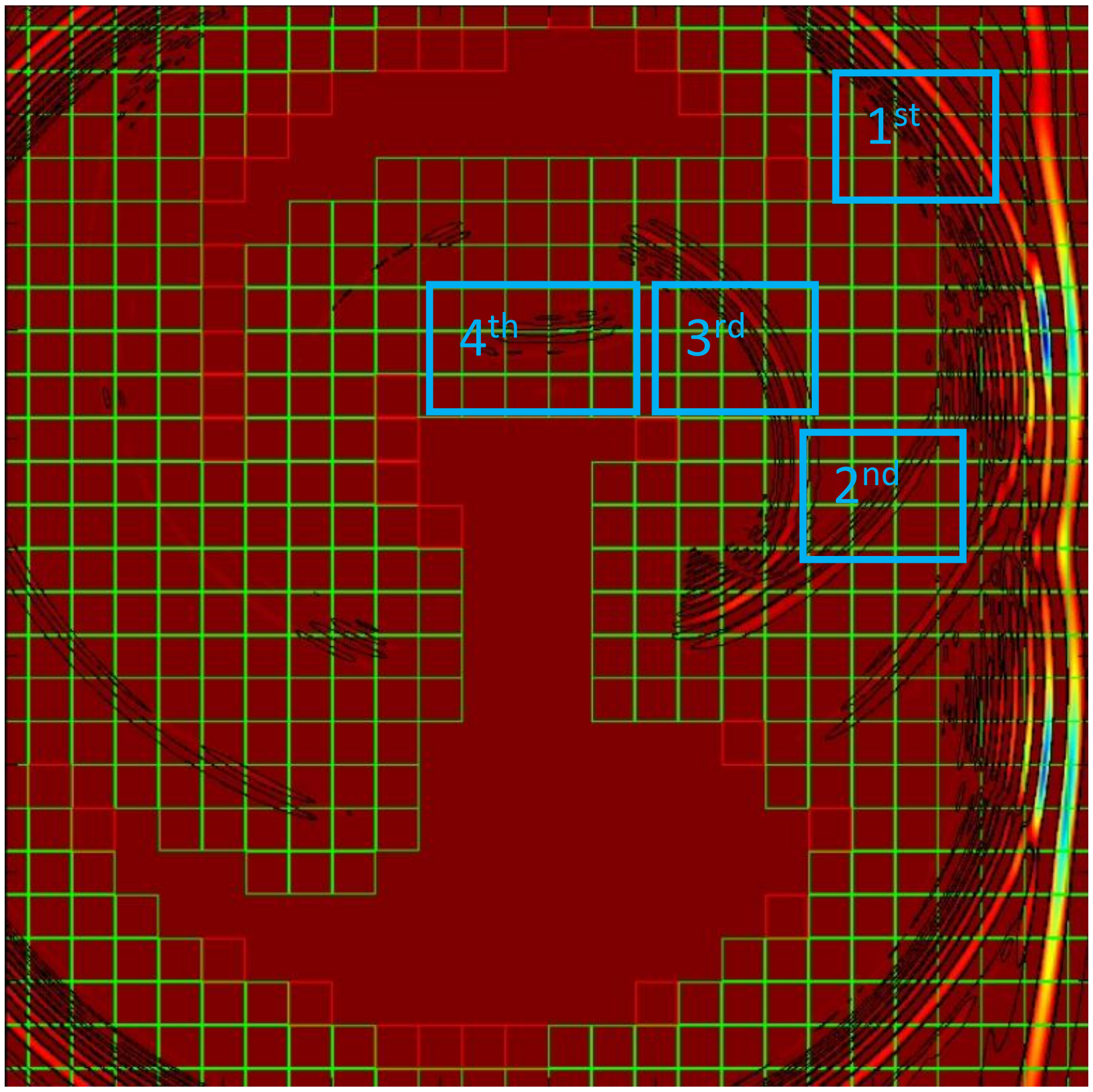

**Figure 13 - The $1^{st}$ to $4^{th}$ order reflection (sphere – cube – sphere - cube) of the incident wave**

## Conclusions

The TCM shows ground breaking achievements in terms of speedup (a factor of *26* before parallelization) and effective memory increase (a factor of *192*). The method was introduced, the mechanism of the hybrid computational engine was explained along with the HBM, for describing sparsely scattering media.

There are numerous parameters that heavily influence the performance of the TCM. Finding the optimal parameters for all the variables seems unnecessary, for each scenario running on each machine will demonstrate slightly different results.

## References

---

[1] K. YEE, “Numerical solution of initial boundary value involving Maxwell’s equations problem in isotropic media”, *IEEE Trans. Antennas Propagat*. Vol. AP-14, pp 302-307, May 1966.
[2] A. Taflove, “Numerical stability”, in “Computational Electromagnetics – The finite difference time domain method”. Artec House, 1995, pp 81-92.
[3] Itay Naeh and Abraham Katzir, "Perfectly correlated phase screen realization using sparse spectrum harmonic augmentation," Appl. Opt. 53, 6168-6174 (2014)

[4] Mur, G., “Absorbing boundary conditions for the finite difference approximation of the time domain electromagnetic field equations”, *IEEE Trans. Electromagnetic Compatibility*, vol. 23, 1981, pp. 377-382.
[5] Berenger, J. –P., “A perfectly mached layer for the absorption of electromagnetic waves”, *J. Computational physics,* vol. 114, 1994, pp. 185-200.